\documentclass[%
 reprint,
 amsmath,amssymb,
 aps,
]{revtex4-2}

\usepackage{graphicx}
\usepackage{dcolumn}
\usepackage{hyperref}
\usepackage{bm}

\begin{document}

\preprint{APS/123-QED}

\title{Validation of an Ab Initio-Informed Electronic Stopping Model\\for Large-Scale Atomistic Simulations}

\author{G. P. Kiely}
 \email{glen.kiely@aalto.fi}
\author{R. Nuñez-Palacio}
\author{A. E. Sand}%
\affiliation{%
 Department of Applied Physics, Aalto University, 00076 Aalto Espoo, Finland
}%

\date{\today}

\begin{abstract}
Accurately modeling nonadiabatic electronic energy dissipation in atomistic simulations of radiation damage remains a key challenge. Existing models often rely on arbitrary thresholds and neglect trajectory-dependent, band-structure, and nonlinear effects. A model developed by Tamm \textit{et al.} (\href{https://doi.org/10.1103/PhysRevLett.120.185501}{doi:10.1103/PhysRevLett.120.185501}) offers a nonempirical description of electronic dissipation for atomistic simulations, but has not been directly benchmarked in the electronic stopping regime. We present the first such benchmark via the comparison of atomistic simulations with experimental ion transmission experiments. The model accurately captures the magnitude and trajectory-dependence of the predicted energy losses. These results support the reliability of the model for predictive large-scale atomistic simulations of radiation cascades.


\end{abstract}



\maketitle

Developments in modern nuclear and space technologies require materials that maintain reliable performance under intense radiation, often at extreme pressures and temperatures. 
The reliable use of materials in such environments requires quantitative predictions of the extent of radiation-induced displacement damage that can alter the structural and physical properties of materials \cite{was_materials}. Empirical studies have shone light on the effects of radiation; however, the fundamental mechanisms of damage occur on the timescale of picoseconds, and thus are inaccessible to experimental observation. Atomistic simulations, in particular molecular dynamics (MD), offer the opportunity to investigate radiation damage processes and make damage predictions under many conditions at relatively low expense. 

The widespread success of classical MD can be attributed to the Born-Oppenheimer approximation \cite{bornoppenheimer}, whereby electrons are assumed to remain in their instantaneous ground state and nonadiabatic electronic effects are neglected. This allows for efficient simulation of atomistic systems, as the dynamics of nuclei can be accurately simulated as classical particles that interact via an interatomic potential \cite{Nordlund_review}. However, in displacement cascades, nonadiabatic electronic effects significantly affect atomistic evolution \cite{Correa2012, diaz_role_thermal, zarkadoula2016}. The implementation of these processes in MD is not trivial. 

In displacement cascades, knock-on atoms dissipate kinetic energy in inelastic collisions with electrons, a phenomenon known as electronic stopping \cite{lindhard_ranges}. It has been widely implemented as an extension to classical MD via a velocity-dependent retarding friction force, whose magnitude is, in general, proportional to the velocity of the energetic ion \cite{lindhard_ranges}. However, this simple velocity dependence cannot capture electronic band structure and other nonlinear effects \cite{lim_2016, quashie_2016}, thus missing the strong trajectory-dependence of electronic stopping \cite{NunezPalacio2025}. Moreover, to avoid unphysical quenching of the system, this friction is only applied to atoms with kinetic energy above a certain cutoff, which has no basis in theory and has historically been determined arbitrarily \cite{Zarkadoula2013, Zhurkin2003, zarkadoula2016, Jakas1985_arb_thresh, Nordlund1998_defectproduction}. 

As the kinetic energy during a displacement cascade dissipates, disparity between heat transport efficiencies results in different transient temperatures of the electronic and atomic subsystems. This temperature imbalance is mediated by the exchange of energy between electrons and atoms via electron-phonon (e–ph) coupling \cite{Nordlund_review}. The effect of e-ph coupling in displacement cascades has been of interest for decades \cite{Flynn1988}. Implementations of e-ph coupling in classical MD employ either a damping force \cite{finnis1991thermal, gao1998effects}, which suffers from the same threshold requirement as electronic stopping, or two‑temperature MD (TTMD), where electrons act as a heat bath coupled to the ions \cite{duffy_ttm_model, duffy_elec_cascades, zarkadoula2014, Zhang_EPH_TTM}.

Earlier studies have scrutinized deficiencies in modeling approaches for nonadiabatic electronic effects \cite{Page2009} and highlighted consequences for radiation damage predictions \cite{race2010treatment}. The choice of arbitrary thresholds for both the e-ph coupling and electronic stopping affects the predictions of the amount of damage \cite{bjorkas_nordlund, Page2009, sand_defects_Se} and the morphology of the resulting defects \cite{sand_defects_Se, Sand2013}. Neglecting nonlinear, band structure and trajectory-dependent effects on electronic stopping results in unphysical ion penetration depths \cite{UTTM_MDRANGE, sand2019, zhenyaranges}. Similarly, a two-temperature treatment affects the energy transferred to the electronic system and on atomistic displacements \cite{uyo_si_UTTM}. These limitations have motivated the development of more accurate approaches.

Caro and Victoria suggested in 1989 that the e-ph coupling is related to the low-velocity limit of electronic stopping \cite{carovic}. With advances in modern computational capabilities and real-time implementations, time-dependent density functional theory (rt-TDDFT) has enabled the accurate first-principles calculation of nonadiabatic electronic effects \cite{correa_2018_esfp}. Caro et al. subsequently revisited the original hypothesis using modern TDDFT methods and demonstrated that the e-ph coupling can indeed be calculated as a particular case of electronic stopping \cite{caro2015adequacy}, which led to the work of Tamm et al. who introduced a model for these effects as friction forces modulated by a scalar damping coefficient \cite{tamm2016electron}. However, this scalar friction treatment indiscriminately damps all collective motion and fails to capture phonon polarization lifetimes. Tamm et al. addressed  this insufficiency by developing a more general framework based on tensorial friction and correlated random fluctuations \cite{tamm_2018_eph2}, and implemented it in a TTMD framework \cite{tamm_2019_eph3, user-eph, LAMMPS}. This model, referred to as the unified two-temperature model (UTTM), simultaneously captures the weak and strong damping of the respective e-ph coupling and electronic stopping regimes \cite{tamm_2019_eph3}. 

In the UTTM, correlated friction and stochastic contributions are introduced alongside the interatomic forces of classical MD via the equation:

\begin{equation} \label{eq:full_EPH}
\mathbf{f}_I = -\nabla_I U - \underbrace{\sum_J \mathbf{B}_{IJ} \mathbf{v}_J}_{\boldsymbol{\sigma}_I} + \underbrace{\sum_J \mathbf{W}_{IJ} \boldsymbol{\xi}_J}_{\boldsymbol{\eta}_I}
\end{equation}

where $\bm{\sigma}_I$ is a tensorial friction, proportional to the relative velocities of the neighborhood atoms, and corresponding random fluctuations are described by $\boldsymbol{\eta}_I$. The $\bm{B}_{IJ} = \sum_K \bm{W}_{IK} \bm{W}^T_{JK}$ tensor is positive-definite, with:

\begin{equation} \label{frictioncalceq}
    \bm{W}_{IJ} =
    \begin{cases}
        -\alpha_J \frac{\rho_I \left( r_{IJ} \right)}{\bar{\rho}_J} \bm{e}_{IJ} \bm{e}_{IJ}, & (I \neq J) \\
        \alpha_I \sum_{K \neq I} \frac{\rho_K \left( r_{IK} \right)}{\bar{\rho}_I} \bm{e}_{IK} \bm{e}_{IK}, & (I = J).
    \end{cases}
\end{equation}

The total electron density at any point $\bar{\rho}$ regulates the damping strength and is calculated locally as the sum of the contributions of individual densities $\rho(r)$ of neighbors. The coupling function $\alpha = \alpha(\bar{\rho})$ uses total electron density $\bar{\rho}$ as a proxy for the magnitude of the local electronic coupling. The random forces $\boldsymbol{\eta}$ are determined by a set of independent white noise Gaussian variables $\langle \xi(t)\rangle$ given by:

\begin{equation} \label{eq:rand_forces}
\langle \xi_I(t) \xi_J(t') \rangle = 2 k_B T_e \delta(t - t') \delta_{IJ}
\end{equation}

where $T_e$ is the local temperature of the electronic system \cite{tamm_2018_eph2}. The model contains no free or empirical parameters, and the $\alpha (\bar{\rho})$ coupling parameter is fit to rt-TDDFT data, offering \textit{ab initio} accuracy of the electronic energy dissipation. Parameterizations for the UTTM have already been developed for a variety of materials \cite{caro_2019_eph1, jarrin_si_coupling, teunissen2023, rafael_elemental, zhenya_preprint_2026}.

Although simulations using the UTTM have shown reasonable results \cite{nickel_AS_GPK, jarrin_si_coupling, teunissen2023, uyo_si_UTTM}, its treatment of electronic energy dissipation has not been directly validated in the electronic stopping regime. In this Letter, we report the first direct benchmark of UTTM-predicted electronic energy loss in this regime by reproducing the ion transmission measurements of Lohmann \textit{et al.} \cite{lohmann_2020_si}. To achieve a direct comparison with experimental measurements, we have implemented the UTTM in MDRANGE \cite{nordlund_1994_mdrange}, an efficient molecular-dynamics-based code for ion-range calculations. Details on the implementation can be found in Ref. \cite{UTTM_MDRANGE}. Using this, we simulate a beam of 100-keV $\mathrm{Si}$ ions transmitting through a 53-nm monocrystalline $\mathrm{Si}$ foil, and measure their exit energies on a circular detector 0.29 m downstream. Electronic energy dissipation is calculated using the UTTM parametrization by Núñez-Palacio \textit{et al.} for silicon \cite{rafael_elemental}. By varying the foil orientation relative to the incident beam, the experiment probes energy lost by ions traveling along distinct crystallographic trajectories, allowing the direct evaluation of the trajectory-dependence of the UTTM-calculated electronic energy dissipation. 

\begin{figure}
\includegraphics[width=0.45\textwidth]{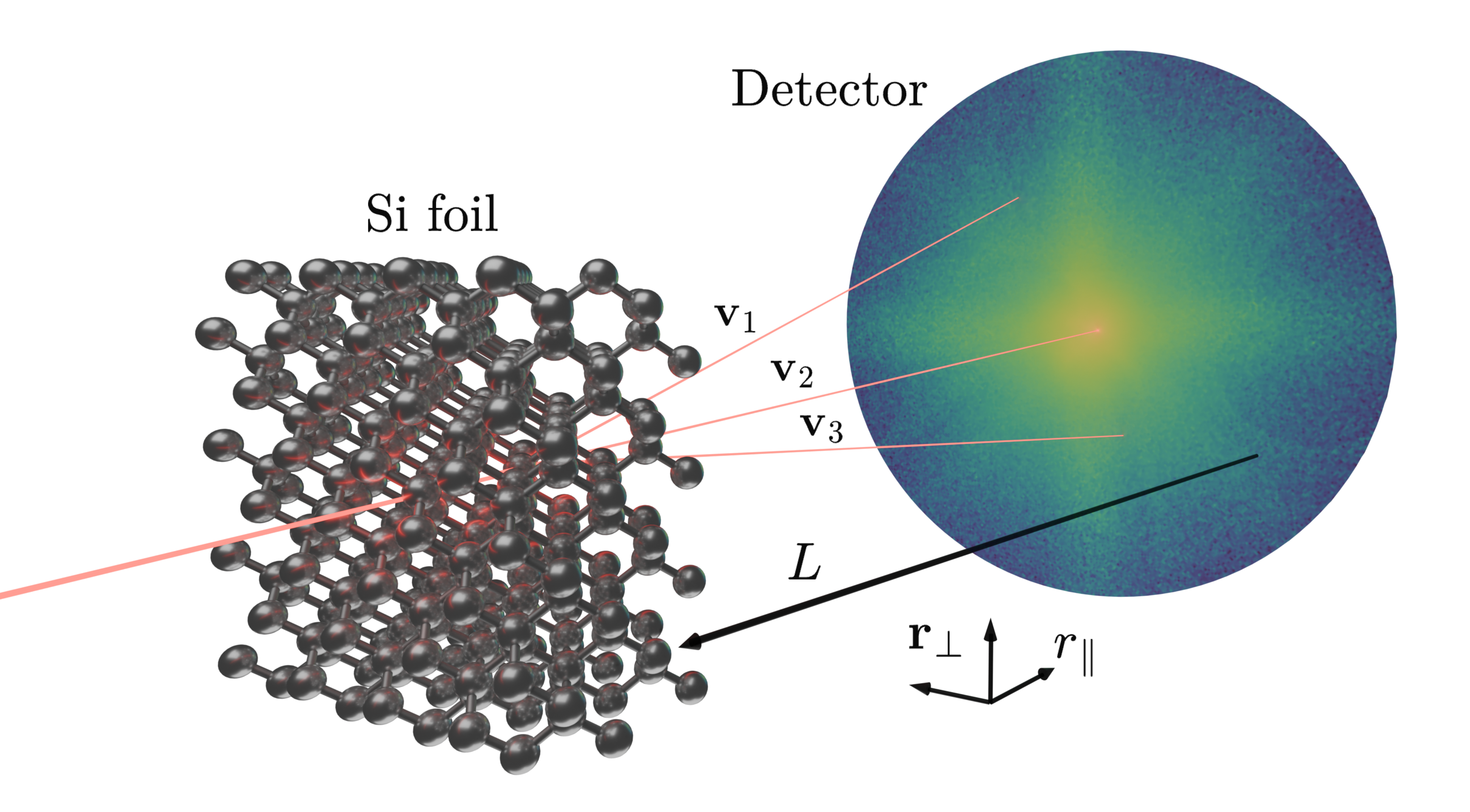}
\caption{\label{fig:schematic_analysis} Schematic of the ion trajectory analysis post-MD simulations. Ions are mapped to a detector plane at a distance $L$ according to their individual exit velocities upon transmission through an Si foil.
}
\end{figure}

We perform otherwise identical simulations using different electronic stopping-power models implemented in MDRANGE to compare with the UTTM predictions. We employ a scalar friction model with velocity-dependent stopping calculated using SRIM \cite{SRIM} or time-dependent density functional theory (TDDFT) \cite{NunezPalacio2025}, and an additional trajectory-dependent model based on local Firsov and free electron-gas theory \cite{peltola2005}. A silicon lattice with a lattice constant of $5.43$ Å was replicated on-the-fly as ions propagated through the foil. The foil was thermalized to 300 K with displacements determined by the Debye model \cite{debyenordlund}. Projectile Si ions were initialized at random positions near the foil surface. Interactions between the projectiles and foil atoms were described by the Ziegler–Biersack–Littmark (ZBL) interatomic potential \cite{ZBL_book}, which accurately describes the short-range nuclear repulsion relevant in high-energy collisions. Each simulation was performed until 25 million ions transmitted through the foil. Beam orientations were varied to reproduce the channeling and pseudo-random directions investigated experimentally \cite{lohmann_2020_si}.

The transmitted ions from MDRANGE were propagated to the detector in post-analysis via: 
\begin{equation}
\mathbf{r}_{\perp,d}
=
\mathbf{r}_{\perp,t}
+
\frac{L}{v_{\parallel}}\mathbf{v}_{\perp},
\end{equation}
where $L=r_{\parallel,d}-r_{\parallel,t}=290~\mathrm{mm}$ is the distance to the detector along the axis normal to both surfaces and parallel to the beam direction.
Here, $\mathbf{r}_{\perp,t}$ and $\mathbf{r}_{\perp,d}$ denote the transverse ion positions at the foil exit and detector, respectively, while $\mathbf{v}_{\perp}$ and $v_{\parallel}$ are the transverse and beam-parallel components of the ion velocity upon exiting the foil. This analysis is illustrated in Fig. \ref{fig:schematic_analysis}.

\begin{figure}
\includegraphics[width=0.49\textwidth]{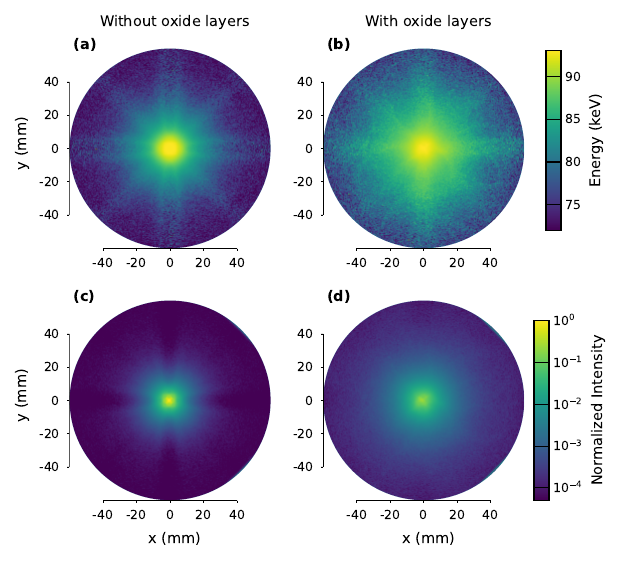}
\caption{\label{fig:foilcomp} Effect of surface oxide layers on simulated transmission of 100 keV $^{29}\mathrm{Si}$ ions through a 53-nm-thick Si foil under $\langle001\rangle$ channeling incidence. Position-dependent mean detected energy (a) without oxide layers and (b) with 22-\AA-thick oxide layers on the entry and exit surfaces. Corresponding detected ion intensity (c) without oxide layers and (d) with oxide layers. Electronic energy dissipation was modeled using the UTTM parametrization for Si developed by Núñez-Palacio \textit{et al.} \cite{rafael_elemental}.
}
\end{figure}

\begin{figure*}
\includegraphics[trim=1cm 0cm 1cm 0cm, clip, width=0.95\textwidth]{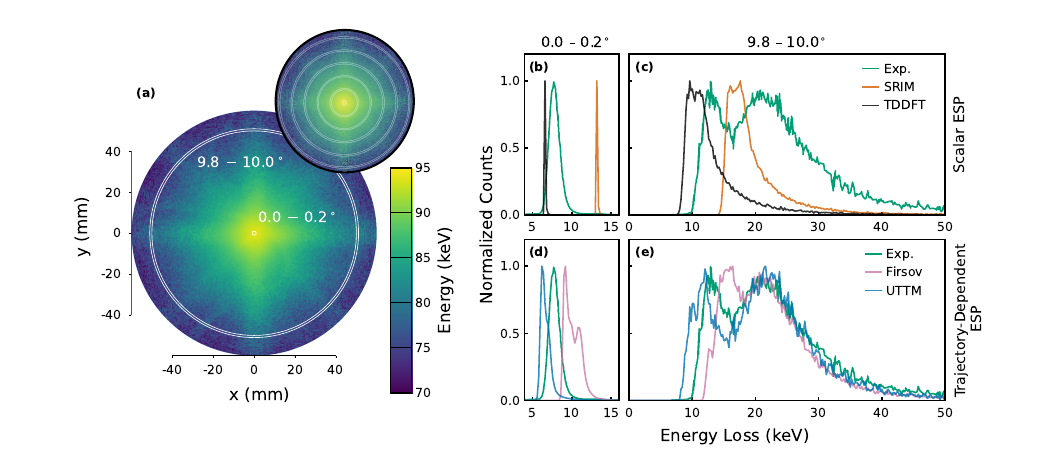}
\caption{\label{fig:lohmann_SRIM_DFT} 
Simulated transmission of 100-keV $^{29}\mathrm{Si}$ ions through a 53-nm-thick Si foil for $\langle001\rangle$ channeling incidence. (a) Map of the mean detected energy. The inset shows the corresponding experimental detector image from Ref. \cite{lohmann_2020_si} on the same color scale. Energy-loss distributions are shown for ions detected in the central scattering annulus, $0.0^\circ$ to $0.2^\circ$, using (b) scalar electronic stopping models and (d) trajectory-dependent models. Corresponding distributions for the outer annulus, $9.8^\circ$ to $10.0^\circ$, are shown using (c) scalar models and (e) trajectory-dependent models. The UTTM parametrization for Si was developed by Núñez-Palacio \textit{et al.} \cite{rafael_elemental}. The experimental image in (a) is reproduced from Ref.\cite{lohmann_2020_si} under the CC BY 4.0 license. 
}
\end{figure*}

To assess the material specifications, we note Lohmann \textit{et al.}'s speculation that the character of the measured ion energy-loss distributions was influenced by the presence of oxide layers on the foil surfaces \cite{lohmann_2020_si}. Holeňák \textit{et al.} subsequently demonstrated that the inclusion of oxide layers is necessary to reproduce experimentally observed energy-loss distributions in BCA simulations \cite{holenak2024}. In agreement with these previous observations, in our simulations 22~\AA-thick oxide layers were indeed required on both foil surfaces to reproduce the degree of scattering observed in the experimental energy-loss maps of Lohmann \textit{et al.} \cite{lohmann_2020_si}.
For $\langle 001 \rangle$ channeling incidence, including oxide layers results in a more uniform distribution of the mean detected energy per pixel (Fig. \ref{fig:foilcomp}a–b) and detected intensity (Fig. \ref{fig:foilcomp}c–d). 
The effect of the oxide layers is to scatter ions at the entrance and exit surfaces, resulting in a more uniform distribution of ions across the detector. A similar effect of scattering was observed in experimental work \cite{Holenak_Contrast} for higher incident beam energies, where a reduced scattering probability results in less uniform detector images.

The energy map of transmitted ions with $\langle001\rangle$ channeling incidence is shown in Fig. \ref{fig:lohmann_SRIM_DFT}(a). There is excellent agreement with the experimental map of Lohmann \textit{et al.} \cite{lohmann_2020_si} shown in the inset. We chose two representative annuli from the experimental work \cite{lohmann_2020_si} to quantify the simulated energy loss predictions: the central transmitted beam, $0.0$--$0.2^\circ$, containing strongly channeled trajectories, and an outer annulus at $9.8$--$10.0^\circ$, which contains ions sampling a broader range of trajectories. In the outer annulus, the characteristic two-peak structure shows the contributions from both channeled and randomly scattered ions \cite{lohmann_2020_si}. Figure \ref{fig:lohmann_SRIM_DFT}(b,c) show the distributions obtained with scalar SRIM and TDDFT-calculated electronic stopping powers, while Fig.\ref{fig:lohmann_SRIM_DFT}(d,e) compare two trajectory-dependent models: the UTTM and the silicon-specific model developed by Peltola \textit{et al.} \cite{peltola2005}.

\begin{figure*}
\includegraphics[trim=1cm 0cm 1cm 0.5cm, clip, width=0.95\textwidth]{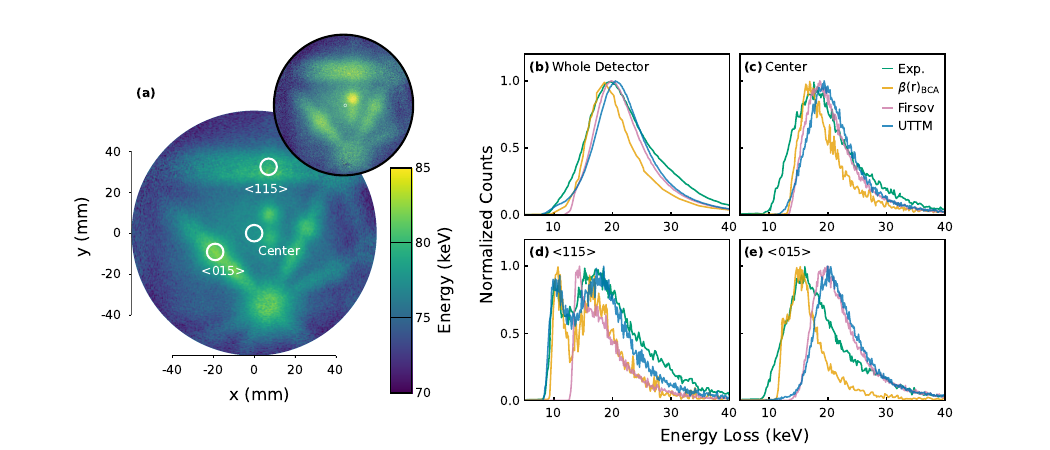}
\quad \quad 
\caption{\label{fig:random_dir_plots} 
Simulated transmission of 100-keV $^{29}\mathrm{Si}$ ions through a 53-nm-thick Si foil for pseudorandom incidence. The foil was rotated by $6^\circ$ about the $x$ axis and $12^\circ$ about the $y$ axis relative to the $\langle001\rangle$ channeling geometry in Fig.\ref{fig:lohmann_SRIM_DFT}. (a) Map of the mean detected energy. The inset shows the corresponding experimental detector image from Ref.\cite{lohmann_2020_si} on the same color scale. Energy-loss distributions are shown for ions detected (b) over the full detector, (c) within a small region around the transmitted-beam position, (d) along the $\langle115\rangle$ direction, and (e) along the $\langle015\rangle$ direction. The experimental image in (a) is reproduced from Ref.~\cite{lohmann_2020_si} under the CC BY 4.0 license.
}
\end{figure*}

Simulations employing scalar models for the electronic stopping power result in much narrower distributions of energy losses than those calculated using trajectory-dependent models. For ions in the central $\langle001\rangle$ channel, the scalar rt-TDDFT stopping yields energy losses close to the experimental central-beam distribution. In contrast, the SRIM stopping results in much larger energy losses. 
The trajectory-dependent models both reproduce the magnitude and the qualitative character of the energy loss distributions more accurately, including wide energy loss distributions and, in the outer annulus, the two-peak structure resulting from channeled and scattered trajectories. 

We further validate the trajectory-dependence of the UTTM using a pseudorandom incidence, whereby the foil is rotated by $6^\circ$ about the $x$ axis and $12^\circ$ about the $y$ axis relative to the $\langle001\rangle$ channeling case \cite{lohmann_2020_si}. Figure\ref{fig:random_dir_plots}(a) shows the simulated mean-energy map of the detected ions, with the corresponding experimental map shown in the inset. The energy-loss distributions integrated over the full detector and over the central beam region are shown in Fig.\ref{fig:random_dir_plots}(b,c), and Fig.\ref{fig:random_dir_plots}(d,e) shows the energy losses corresponding to ions traveling along the $\langle115\rangle$ and $\langle015\rangle$ lattice directions.

For comparison, we also include the energy losses of pseudorandom incident Si ions simulated by Holeñák \textit{et al.} \cite{holenak2024}, who used a local electronic stopping model based on the form proposed by Oen and Robinson \cite{oenrobinson} implemented in a BCA code. While this model results in excellent agreement of energy losses for the presented trajectories, its parameters were fitted empirically for the pseudorandom orientation and are not valid, for example, for $\langle001\rangle$ channeling incidence.
In contrast, the UTTM model is parameter-free, informed by \textit{ab initio} simulations, and is applicable for all of the trajectories considered. We observe good agreement with experiment for most trajectories using the UTTM. In particular, the model reproduces the two-peak structure observed both in the outer annulus for $\langle001\rangle$ incidence, Fig.\ref{fig:lohmann_SRIM_DFT}(e), and in the annulus corresponding to the $\langle115\rangle$ channel for pseudorandom incidence \ref{fig:random_dir_plots}(d). Because the relative heights of these peaks are sensitive to the oxide-layer thickness \cite{holenak2024}, the agreement in both cases supports the use of 22 Å-thick oxide layers in the simulations. For detected ions corresponding to the $\langle015\rangle$ direction, Fig.~\ref{fig:random_dir_plots}(e), the UTTM predicts energy losses that differ in magnitude from experiment. Nevertheless, the widths of the distributions remain in good agreement across all directions.

\begin{figure}
\includegraphics[width=0.49\textwidth]{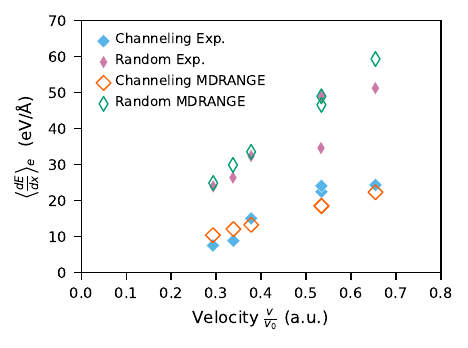}
\quad \quad 
\caption{\label{fig:estopplot} 
Comparison of experimentally measured electronic stopping powers from Ref.~\cite{lohmann_2020_si} with the average effective electronic stopping powers obtained in this work. For each data point, the simulation conditions correspond directly to those of the experiment. Two datapoints overlap for the Channeling MDRANGE case at $\sim 0.53$ a.u.
}
\end{figure}

Moreover, because nuclear stopping is treated explicitly using the repulsive ZBL interatomic potential \cite{ZBL_book} in our simulations, we can directly separate the nuclear and electronic contributions to the total ion energy loss, and hence assess only the electronic stopping contribution along these trajectories. In Fig.~\ref{fig:estopplot}, we compare the effective electronic stopping power extracted from our simulations with the values determined experimentally by Lohmann \textit{et al.} \cite{lohmann_2020_si}. The comparison considers ions detected within the central $0.0^\circ$--$0.2^\circ$ scattering annulus for the case of channeling incidence. Si isotopes and foil thicknesses were varied in our simulations to match the experimental conditions. The average effective electronic stopping power is calculated by:
\begin{equation}
\left\langle \frac{dE}{dx} \right\rangle_e
= \frac{1}{N} \sum_{i=1}^{N} \left( \frac{\Delta E_i^{e}}{d} \right) 
\end{equation}
where $\Delta E_i^{e}$ is the energy lost via electronic dissipation by projectile ion $i$, $d$ is the foil thickness, and $N$ is the number of detected ions included in the average. Across the investigated velocities, the effective electronic stopping power predicted using the UTTM agrees well with the experiment, for both channeling and random incidence.

In conclusion, we demonstrate that the UTTM model for the electronic stopping results in predicted ion energy losses that are in good agreement with experimental results over a range of crystallographic directions. We observe this agreement in both the magnitude and the character of simulated energy losses along all of the investigated directions. The energy loss predictions using the UTTM model show significant improvements compared to the predictions made using other state-of-the-art electronic stopping models. This work serves as the first direct benchmark of the energy dissipation predicted by the UTTM model in the electronic stopping regime. 

\quad

This work was funded by the European Research Council (ERC, MUST, 101077454). Views and opinions expressed are, however, those of the author(s) only and do not necessarily reflect those of the European Union or the European Research Council. Neither the European Union nor the granting authority can be held responsible for them. The authors would like to acknowledge the computational resources provided by the Aalto Science-IT project and thank Radek Holeñák for valuable discussions. 


\bibliography{apssamp}

@book{ZBL_book,
  title={The Stopping and Range of Ions in Solids},
  author={Ziegler, J.F. and Biersack, J.P. and Littmark, U.},
  isbn={9780080216034},
  lccn={85012157},
  series={Stopping and ranges of ions in matter},
  year={1985},
  publisher={Pergamon}
}

@article{zhenyaranges,
title = {Ab initio informed electronic stopping models for light ion propagation in metals},
journal = {J. Nucl. Mater.},
volume = {632},
pages = {156906},
year = {2026},
issn = {0022-3115},
doi = {https://doi.org/10.1016/j.jnucmat.2026.156906},
url = {https://www.sciencedirect.com/science/article/pii/S002231152600471X},
author = {Evgeniia Ponomareva and Artur Tamm and Andrea E. Sand}
}

@article{Zhang_EPH_TTM,
doi = {10.1088/0953-8984/25/23/235402},
year = {2013},
month = {may},
publisher = {IOP Publishing},
volume = {25},
number = {23},
pages = {235402},
author = {Zhang, Chao and Mao, Fei and Zhang, Feng-Shou},
title = {Electron–ion coupling effects on radiation damage in cubic silicon carbide},
journal = {J. Phys.: Condens. Matter}
}

@article{bjorkas_nordlund,
title = {Assessment of the relation between ion beam mixing, electron–phonon coupling and damage production in Fe},
journal = {Nucl. Instrum. Methods Phys. Res. B},
volume = {267},
number = {10},
pages = {1830-1836},
year = {2009},
issn = {0168-583X},
doi = {https://doi.org/10.1016/j.nimb.2009.03.080},
author = {C. Björkas and K. Nordlund}
}

@article{duffy_ttm_model,
  title={Including the effects of electronic stopping and electron-ion interactions in radiation damage simulations},
  author={Duffy, DM and Rutherford, AM},
  journal={J. Phys.: Condens. Matter},
  volume={19},
  number={1},
  pages={016207},
  year={2007}
}

@article{duffy_elec_cascades,
title = {Including electronic effects in damage cascade simulations},
journal = {J. Nucl. Mater.},
volume = {386-388},
pages = {19-21},
year = {2009},
issn = {0022-3115},
author = {Duffy, D. and Rutherford, A.},
doi = {https://doi.org/10.1016/j.jnucmat.2008.12.051}
}

@article{carovic,
  title = {Ion-electron interaction in molecular-dynamics cascades},
  author = {Caro, A. and Victoria, M.},
  journal = {Phys. Rev. A},
  volume = {40},
  issue = {5},
  pages = {2287--2291},
  numpages = {0},
  year = {1989},
  month = {Sep},
  publisher = {American Physical Society},
  doi = {10.1103/PhysRevA.40.2287},
  url = {https://link.aps.org/doi/10.1103/PhysRevA.40.2287}
}

@article{finnis1991thermal,
  title={Thermal excitation of electrons in energetic displacement cascades},
  author={Finnis, MW and Agnew, P and Foreman, AJE},
  journal={Phy. Rev. B},
  volume={44},
  number={2},
  pages={567},
  year={1991},
  publisher={APS}
}

@article{oenrobinson,
title = {Computer studies of the reflection of light ions from solids},
journal = {Nucl. Instrum. Methods},
volume = {132},
pages = {647-653},
year = {1976},
issn = {0029-554X},
doi = {https://doi.org/10.1016/0029-554X(76)90806-5},
url = {https://www.sciencedirect.com/science/article/pii/0029554X76908065},
author = {Ordean {S. Oen} and Mark {T. Robinson}}
}

@article{Nordlund_review,
title = {Primary radiation damage: A review of current understanding and models},
journal = {J. Nucl. Mater.},
volume = {512},
pages = {450-479},
year = {2018},
issn = {0022-3115},
doi = {https://doi.org/10.1016/j.jnucmat.2018.10.027},
author = {Kai Nordlund and Steven J. Zinkle and Andrea E. Sand and Fredric Granberg and Robert S. Averback and Roger E. Stoller and Tomoaki Suzudo and Lorenzo Malerba and Florian Banhart and William J. Weber and Francois Willaime and Sergei L. Dudarev and David Simeone}
}

@misc{rafael_elemental,
  author  = {Nuñez-Palacio, Rafael and Kiely, Glen and Tamm, Artur and Sand, Andrea},
  title   = {Trajectory-Dependent Electronic Stopping in Simulations of Self-Ion Ranges in Elemental Semiconductors},
  year    = {2026},
  month   = {August},
  journal = {Research Square},
  note    = {Preprint},
  doi     = {10.21203/rs.3.rs-9471547/v1},
  url     = {https://doi.org/10.21203/rs.3.rs-9471547/v1}
}

@misc{zhenya_preprint_2026,
      title={Modeling anisotropic energy dissipation of light ions at the atomistic scale}, 
      author={Evgeniia Ponomareva and Artur Tamm and Andrea E. Sand},
      year={2026},
      eprint={2603.10838},
      archivePrefix={arXiv},
      primaryClass={cond-mat.mtrl-sci},
}

@article{Holenak_Contrast,
title = {Contrast modes in a 3D ion transmission approach at keV energies},
journal = {Ultramicroscopy},
volume = {217},
pages = {113051},
year = {2020},
issn = {0304-3991},
doi = {https://doi.org/10.1016/j.ultramic.2020.113051},
author = {R. Holeňák and S. Lohmann and D. Primetzhofer}
}

@article{caro2015adequacy,
  title={Adequacy of damped dynamics to represent the electron-phonon interaction in solids},
  author={Caro, Alfredo and Correa, AA and Tamm, A and Samolyuk, German D and Stocks, George M},
  journal={Phys. Rev. B},
  volume={92},
  number={14},
  pages={144309},
  year={2015},
  publisher={APS}
}

@article{tamm2016electron,
  title={Electron-phonon interaction within classical molecular dynamics},
  author={Tamm, A and Samolyuk, G and Correa, AA and Klintenberg, Mattias and Aabloo, A and Caro, A},
  journal={Phys. Rev. B},
  volume={94},
  number={2},
  pages={024305},
  year={2016},
  publisher={APS}
}

@article{was_materials,
title = {Materials challenges in nuclear energy},
journal = {Acta Mater.},
volume = {61},
number = {3},
pages = {735-758},
year = {2013},
issn = {1359-6454},
doi = {https://doi.org/10.1016/j.actamat.2012.11.004},
author = {S.J. Zinkle and G.S. Was}
}

@article{debyenordlund,
title = {Dependence of ion channeling on relative atomic number in compounds},
journal = {Nucl. Instrum. Methods Phys. Res. B},
volume = {435},
pages = {61-69},
month = {Nov},
year = {2018},
issn = {0168-583X},
doi = {https://doi.org/10.1016/j.nimb.2017.11.020},
author = {K. Nordlund and G. Hobler}
}

@Article{LAMMPS,
  author = "A. P. Thompson and H. M. Aktulga and R. Berger and 
     D. S. Bolintineanu and W. M. Brown and P. S. Crozier and
     P. J. in 't Veld and A. Kohlmeyer and S. G. Moore and T. D. Nguyen and
     R. Shan and M. J. Stevens and J. Tranchida and C. Trott and S. J. Plimpton",
  title = "{LAMMPS} - a flexible simulation tool for
     particle-based materials modeling at the 
     atomic, meso, and continuum scales",
  journal = "Comp. Phys. Comm.",
  volume =  "271",
  pages =   "108171",
  year =    "2022",
  doi = "10.1016/j.cpc.2021.108171"
}

@article{bornoppenheimer,
author = {Born, M. and Oppenheimer, R.},
title = {Zur Quantentheorie der Molekeln},
journal = {Ann. Phys. (Berlin)},
volume = {389},
number = {20},
pages = {457-484},
doi = {https://doi.org/10.1002/andp.19273892002},
year = {1927}
}

@misc{user-eph,
  note = {{USER-EPH LAMMPS Fix Codebase:} https://github.com/LLNL/USER-EPH},
  year = {}
}

@article{sand_defects_Se,
title = {On the lower energy limit of electronic stopping in simulated collision cascades in Ni, Pd and Pt},
journal = {J. Nucl. Mater.},
volume = {456},
pages = {99-105},
year = {2015},
issn = {0022-3115},
doi = {https://doi.org/10.1016/j.jnucmat.2014.09.029},
author = {A.E. Sand and K. Nordlund}
}

@article{gao1998effects,
  title={The effects of electron-phonon coupling on defect production by displacement cascades in iron},
  author={Gao, F and Bacon, DJ and Flewitt, PEJ and Lewis, TA},
  journal={Model. Simul. Mater. Sci. Eng.},
  volume={6},
  number={5},
  pages={543--556},
  year={1998}
}

@article{teunissen2023,
  title = {Effect of electronic stopping in molecular dynamics simulations of collision cascades in gallium arsenide},
  author = {Teunissen, Johannes L. and Jarrin, Thomas and Richard, Nicolas and Koval, Natalia E. and Santiburcio, Daniel Mu\~noz and Kohanoff, Jorge and Artacho, Emilio and Cleri, Fabrizio and Da Pieve, Fabiana},
  journal = {Phys. Rev. Mater.},
  volume = {7},
  issue = {2},
  pages = {025404},
  numpages = {13},
  year = {2023},
  month = {Feb},
  publisher = {American Physical Society},
  doi = {10.1103/PhysRevMaterials.7.025404}
}

@article{uyo_si_UTTM,
title = {Improved capabilities of the TurboGAP code for radiation induced cascade simulations: An illustration with silicon},
journal = {Comput. Mater. Sci.},
volume = {267},
pages = {114560},
year = {2026},
issn = {0927-0256},
doi = {https://doi.org/10.1016/j.commatsci.2026.114560},
url = {https://www.sciencedirect.com/science/article/pii/S0927025626000790},
author = {Uttiyoarnab Saha and Ali Hamedani and Miguel A. Caro and Andrea E. Sand}
}

@article{correa_2018_esfp,
title = {Calculating electronic stopping power in materials from first principles},
journal = {Comput. Mater. Sci.},
volume = {150},
pages = {291-303},
year = {2018},
issn = {0927-0256},
doi = {https://doi.org/10.1016/j.commatsci.2018.03.064},
author = {Alfredo A. Correa}
}

@article{Jakas1985_arb_thresh,
  title = {Dependence of atom ejection on electronic energy loss},
  author = {Jakas, Mario M. and Harrison, Don E.},
  journal = {Phys. Rev. B},
  volume = {32},
  issue = {5},
  pages = {2752--2760},
  numpages = {0},
  year = {1985},
  month = {Sep},
  publisher = {American Physical Society},
  doi = {10.1103/PhysRevB.32.2752},
  url = {https://link.aps.org/doi/10.1103/PhysRevB.32.2752}
}

@article{Zarkadoula2013,
doi = {10.1088/0953-8984/25/12/125402},
year = {2013},
month = {feb},
publisher = {IOP Publishing},
volume = {25},
number = {12},
pages = {125402},
author = {Zarkadoula, E and Daraszewicz, S L and Duffy, D M and Seaton, M A and Todorov, I T and Nordlund, K and Dove, M T and Trachenko, K},
title = {The nature of high-energy radiation damage in iron},
journal = {J. Phys.: Condens. Matter}
}

@article{race2010treatment,
  title={The treatment of electronic excitations in atomistic models of radiation damage in metals},
  author={Race, CP and Mason, DR and Finnis, MW and Foulkes, WMC and Horsfield, AP and Sutton, AP},
  journal={Rep. Prog. Phys.},
  volume={73},
  number={11},
  pages={116501},
  year={2010}
}

@article{Nordlund1998_defectproduction,
  title = {Defect production in collision cascades in elemental semiconductors and fcc metals},
  author = {Nordlund, K. and Ghaly, M. and Averback, R. S. and Caturla, M. and Diaz de la Rubia, T. and Tarus, J.},
  journal = {Phys. Rev. B},
  volume = {57},
  issue = {13},
  pages = {7556--7570},
  numpages = {0},
  year = {1998},
  month = {Apr},
  publisher = {American Physical Society},
  doi = {10.1103/PhysRevB.57.7556},
  url = {https://link.aps.org/doi/10.1103/PhysRevB.57.7556}
}

@article{Page2009,
  title={How good is damped molecular dynamics as a method to simulate radiation damage in metals?},
  author={Le Page, J and Mason, DR and Race, CP and Foulkes, WMC},
  journal = {New J. Phys.},
  volume={11},
  number={1},
  pages={013004},
  year={2009}
}

@article{Correa2012,
  title = {Nonadiabatic Forces in Ion-Solid Interactions: The Initial Stages of Radiation Damage},
  author = {Correa, Alfredo A. and Kohanoff, Jorge and Artacho, Emilio and S\'anchez-Portal, Daniel and Caro, Alfredo},
  journal = {Phys. Rev. Lett.},
  volume = {108},
  issue = {21},
  pages = {213201},
  numpages = {5},
  year = {2012},
  month = {May},
  publisher = {American Physical Society},
  doi = {10.1103/PhysRevLett.108.213201}
}

@article{zarkadoula2016,
title = {Effects of two-temperature model on cascade evolution in Ni and NiFe},
journal = {Scripta Mater.},
volume = {124},
pages = {6-10},
year = {2016},
issn = {1359-6462},
doi = {https://doi.org/10.1016/j.scriptamat.2016.06.028},
author = {Eva Zarkadoula and German Samolyuk and Haizhou Xue and Hongbin Bei and William J. Weber}
}

@misc{SRIM,
title = {SRIM Software Package},
url = {https://www.srim.org},
author = {James F. Ziegler},
year = {}
}

@article{caro_2019_eph1,
  title = {Role of electrons in collision cascades in solids. I. Dissipative model},
  author = {Caro, M. and Tamm, A. and Correa, A. A. and Caro, A.},
  journal = {Phys. Rev. B},
  volume = {99},
  issue = {17},
  pages = {174301},
  numpages = {9},
  year = {2019},
  month = {May},
  publisher = {American Physical Society},
  doi = {10.1103/PhysRevB.99.174301}
}

@article{Zhurkin2003,
title = {Atomic scale modelling of Al and Ni(111) surface erosion under cluster impact},
journal =  {Nucl. Instrum. Methods Phys. Res. B},
volume = {202},
pages = {269-277},
year = {2003},
issn = {0168-583X},
doi = {https://doi.org/10.1016/S0168-583X(02)01868-2},
author = {Eugeni E Zhurkin and Anton S Kolesnikov}
}

@article{tamm_2018_eph2,
  title = {Langevin Dynamics with Spatial Correlations as a Model for Electron-Phonon Coupling},
  author = {Tamm, A. and Caro, M. and Caro, A. and Samolyuk, G. and Klintenberg, M. and Correa, A. A.},
  journal = {Phys. Rev. Lett.},
  volume = {120},
  issue = {18},
  pages = {185501},
  numpages = {6},
  year = {2018},
  month = {May},
  publisher = {American Physical Society},
  doi = {10.1103/PhysRevLett.120.185501}
}

@article{zarkadoula2014,
  title={Electronic effects in high-energy radiation damage in iron},
  author={Zarkadoula, Eva and Daraszewicz, SL and Duffy, Dorothy M and Seaton, MA and Todorov, Ilian T and Nordlund, Kai and Dove, Martin T and Trachenko, Kostya},
  journal={J. Phys.: Condens. Matter},
  volume={26},
  number={8},
  pages={085401},
  year={2014},
  publisher={IOP Publishing}
}

@article{Sand2013,
doi = {10.1209/0295-5075/103/46003},
url = {https://doi.org/10.1209/0295-5075/103/46003},
year = {2013},
month = {sep},
publisher = {EDP Sciences, IOP Publishing and Società Italiana di Fisica},
volume = {103},
number = {4},
pages = {46003},
author = {Sand, A. E. and Dudarev, S. L. and Nordlund, K.},
title = {High-energy collision cascades in tungsten: Dislocation loops structure and clustering scaling laws},
journal = {Europhys. Lett.}
}

@article{Flynn1988,
  title = {Electron-phonon interactions in energetic displacement cascades},
  author = {Flynn, C. P. and Averback, R. S.},
  journal = {Phys. Rev. B},
  volume = {38},
  issue = {10},
  pages = {7118(R)--7120(R)},
  numpages = {0},
  year = {1988},
  month = {Oct},
  publisher = {American Physical Society},
  doi = {10.1103/PhysRevB.38.7118},
  url = {https://link.aps.org/doi/10.1103/PhysRevB.38.7118}
}

@article{tamm_2019_eph3,
  title = {Role of electrons in collision cascades in solids. II. Molecular dynamics},
  author = {Tamm, A. and Caro, M. and Caro, A. and Correa, A. A.},
  journal = {Phys. Rev. B},
  volume = {99},
  issue = {17},
  pages = {174302},
  numpages = {9},
  year = {2019},
  month = {May},
  publisher = {American Physical Society},
  doi = {10.1103/PhysRevB.99.174302}
}

@article{NunezPalacio2025,
  author    = {Nu{\~n}ez-Palacio, Rafael and Sand, Andrea E.},
  title     = {Path-dependent electronic stopping for self-irradiated silicon},
  journal   = {Commun. Mater.},
  year      = {2025},
  volume    = {6},
  number    = {1},
  pages     = {109},
  doi       = {10.1038/s43246-025-00834-y},
  issn      = {2662-4443}
}

@article{UTTM_MDRANGE,
title = {Trajectory-dependent electronic energy losses in ion range simulations},
journal = {Comput. Mater. Sci.},
volume = {263},
pages = {114451},
year = {2026},
issn = {0927-0256},
doi = {https://doi.org/10.1016/j.commatsci.2025.114451},
author = {Glen P. Kiely and Bruno Semião and Evgeniia Ponomareva and Rafael Nuñez-Palacio and Unna Arpiainen and Andrea E. Sand}
}

@article{nickel_AS_GPK,
  title = {Electronic effects in radiation-induced collision cascades in nickel},
  author = {Sand, Andrea E. and Kiely, Glen P. and Tamm, Artur and Correa, Alfredo A.},
  journal = {Phys. Rev. Mater.},
  volume = {9},
  issue = {5},
  pages = {053603},
  numpages = {13},
  year = {2025},
  month = {May},
  publisher = {American Physical Society},
  doi = {10.1103/PhysRevMaterials.9.053603}
}

@article{lohmann_2020_si,
  title = {Trajectory-dependent electronic excitations by light and heavy ions around and below the Bohr velocity},
  author = {Lohmann, S. and Hole\ifmmode \check{n}\else \v{n}\fi{}\'ak, R. and Primetzhofer, D.},
  journal = {Phys. Rev. A},
  volume = {102},
  issue = {6},
  pages = {062803},
  numpages = {8},
  year = {2020},
  month = {Dec},
  publisher = {American Physical Society},
  doi = {10.1103/PhysRevA.102.062803}
}

@article{nordlund_1994_mdrange,
title = {Molecular dynamics simulation of ion ranges in the 1–100 keV energy range},
journal = {Comput. Mater. Sci.},
volume = {3},
number = {4},
pages = {448-456},
year = {1995},
issn = {0927-0256},
doi = {https://doi.org/10.1016/0927-0256(94)00085-Q},
author = {K. Nordlund}
}

@article{lim_2016,
  title = {Electron Elevator: Excitations across the Band Gap via a Dynamical Gap State},
  author = {Lim, A. and Foulkes, W. M. C. and Horsfield, A. P. and Mason, D. R. and Schleife, A. and Draeger, E. W. and Correa, A. A.},
  journal = {Phys. Rev. Lett.},
  volume = {116},
  issue = {4},
  pages = {043201},
  numpages = {6},
  year = {2016},
  month = {Jan},
  publisher = {American Physical Society},
  doi = {10.1103/PhysRevLett.116.043201}
}

@article{peltola2005,
author = {J. Peltola, K. Nordlund and J. Keinonen},
title = {Electronic stopping power calculation method for molecular dynamics simulations using local Firsov and free electron-gas models},
journal = {Radiat. Eff. Defects Solids},
volume = {161},
number = {9},
pages = {511--521},
year = {2006},
publisher = {Taylor \& Francis}
}

@article{sand2019,
  author = {Andrea E. Sand and Rafi Ullah and Alfredo A. Correa},
  title = {Heavy ion ranges from first-principles electron dynamics},
  journal = {npj Comput. Mater.},
  month = {Apr},
  year = {2019},
  volume = {5},
  number = {1},
  pages = {43},
  doi = {10.1038/s41524-019-0180-5},
  issn = {2057-3960}
}

@article{jarrin_si_coupling,
  title = {Integration of electronic effects into molecular dynamics simulations of collision cascades in silicon from first-principles calculations},
  author = {Jarrin, Thomas and Richard, Nicolas and Teunissen, Johannes and Da Pieve, Fabiana and H\'emeryck, Anne},
  journal = {Phys. Rev. B},
  volume = {104},
  issue = {19},
  pages = {195203},
  numpages = {15},
  year = {2021},
  month = {Nov},
  publisher = {American Physical Society},
  doi = {10.1103/PhysRevB.104.195203}
}

@article{holenak2024,
  title = {Assessing trajectory-dependent electronic energy loss of keV ions by a binary collision approximation code},
  author = {Hole\ifmmode \check{n}\else \v{n}\fi{}\'ak, R. and Ntemou, E. and Lohmann, S. and Linnarsson, M. and Primetzhofer, D.},
  journal = {Phys. Rev. Appl.},
  volume = {21},
  issue = {2},
  pages = {024048},
  numpages = {11},
  year = {2024},
  month = {Feb},
  publisher = {American Physical Society},
  doi = {10.1103/PhysRevApplied.21.024048}
}

@article{diaz_role_thermal,
  title = {Role of thermal spikes in energetic displacement cascades},
  author = {de la Rubia, T. Diaz and Averback, R. S. and Benedek, R. and King, W. E.},
  journal = {Phys. Rev. Lett.},
  volume = {59},
  issue = {17},
  pages = {1930--1933},
  numpages = {0},
  year = {1987},
  month = {Oct},
  publisher = {American Physical Society},
  doi = {10.1103/PhysRevLett.59.1930},
  url = {https://link.aps.org/doi/10.1103/PhysRevLett.59.1930}
}

@article{quashie_2016,
  title = {Electronic band structure effects in the stopping of protons in copper},
  author = {Quashie, Edwin E. and Saha, Bidhan C. and Correa, Alfredo A.},
  journal = {Phys. Rev. B},
  volume = {94},
  issue = {15},
  pages = {155403},
  numpages = {7},
  year = {2016},
  month = {Oct},
  publisher = {American Physical Society},
  doi = {10.1103/PhysRevB.94.155403}
}

@article{lindhard_ranges,
  author       = {Lindhard, J and Scharff, M and Schioett, H E},
  title        = {Range Concepts and Heavy Ion Ranges},
  url          = {https://www.osti.gov/biblio/4153115},
  journal      = {Kgl. Danske Videnskab. Selskab. Mat. Fys. Medd.  },
  volume       = {Vol: 33: No. 14},
  year         = {1963},
  month        = {01}}

\end{document}